\documentclass[osajnl,twocolumn,showpacs,superscriptaddress,10pt]{revtex4-1} 
\newcommand{\mic}{~\mu \rm m}
\usepackage{amsmath,amssymb,graphicx}
\begin{document}
\title{Energetic mid-IR femtosecond pulse generation by self-defocusing soliton-induced dispersive waves in a bulk quadratic nonlinear crystal
}

\author{B.B. Zhou, H.R. Guo, M. Bache}

\affiliation{DTU Fotonik, Department of Photonics Engineering,
Technical University of Denmark, DK-2800 Kgs. Lyngby, Denmark}




\begin{abstract}
    Generating energetic femtosecond mid-IR pulses is crucial for ultrafast spectroscopy, and currently relies on parametric processes that, while efficient, are also complex.
    Here we experimentally show a simple alternative that uses a single pump wavelength without any pump synchronization and without critical phase-matching requirements. Pumping a bulk quadratic nonlinear crystal (unpoled LiNbO$_3$ cut for noncritical phase-mismatched interaction) with sub-mJ near-IR 50-fs pulses, tunable and broadband ($\sim 1,000$ cm$^{-1}$) mid-IR pulses around $3.0\mic$  are generated with excellent spatio-temporal pulse quality, having up to 10.5 $\mu$J energy (6.3\% conversion). 
    The mid-IR pulses are dispersive waves phase-matched to near-IR self-defocusing solitons created by the induced self-defocusing cascaded nonlinearity. This process is filament-free and the input pulse energy can therefore be scaled arbitrarily by using large-aperture crystals. The technique can readily be implemented with other crystals and laser wavelengths, and can therefore potentially replace current ultrafast frequency-conversion processes to the mid-IR. 
\end{abstract}


\maketitle

Intense ultrashort mid-IR (MIR) pulses with microjoule pulse energy are highly important for the study of molecular vibrations \cite{2005Cowan,2013Ramasesha}, in particular in the 2,800-4,000 cm$^{-1}$ region ($\lambda=2.5-3.6~\mu\rm m$) as it contains the stretching modes of C-H, N-H and O-H bonds. The commonly used methods for generating high-energy ultrashort mid-IR pulses are based on nonlinear optical parametric processes. One of the most popular methods is optical parametric amplification (OPA) \cite{1994Seifert}, and an extra difference-frequency generation (DFG) stage is often integrated to further extend the output wavelength range \cite{2000Kaindl}. These wavelength down-conversion techniques enable the generation of tunable mid-IR pulses in a broad spectral range and have good conversion efficiencies. The downside is that the processes require critical phase-matching condition and synchronized pump wavelengths, which complicates the setup for MIR generation. Four-wave mixing in $\chi^{(3)}$ media has also been reported for intense MIR pulse generation \cite{2001Nienhuys,2007Fuji,2010Petersen}, which also involve synchronized pump wavelengths, and typically bear rather low conversion efficiencies.

\begin{figure}[tb]
\centering
\fbox{  \includegraphics[width=\linewidth]{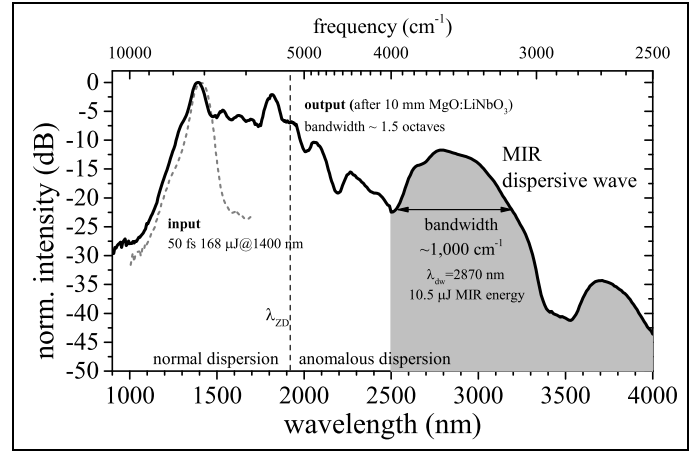}}
  \caption{Normalized input and output spectra from a 10 mm bulk LN crystal pumped with 50 fs 168 $\mu$J pulses centered at $1.4\mic$ and having $0.8~{\rm TW/cm}^2$ peak intensity. A filament-free octave-spanning supercontinuum is formed, including an energetic, broadband DW in the MIR around $2.87\mic$. 
  }\label{fig:spectra}
\end{figure}

\begin{figure}[tb]
  \centering
\fbox{  \includegraphics[width=0.9\linewidth]{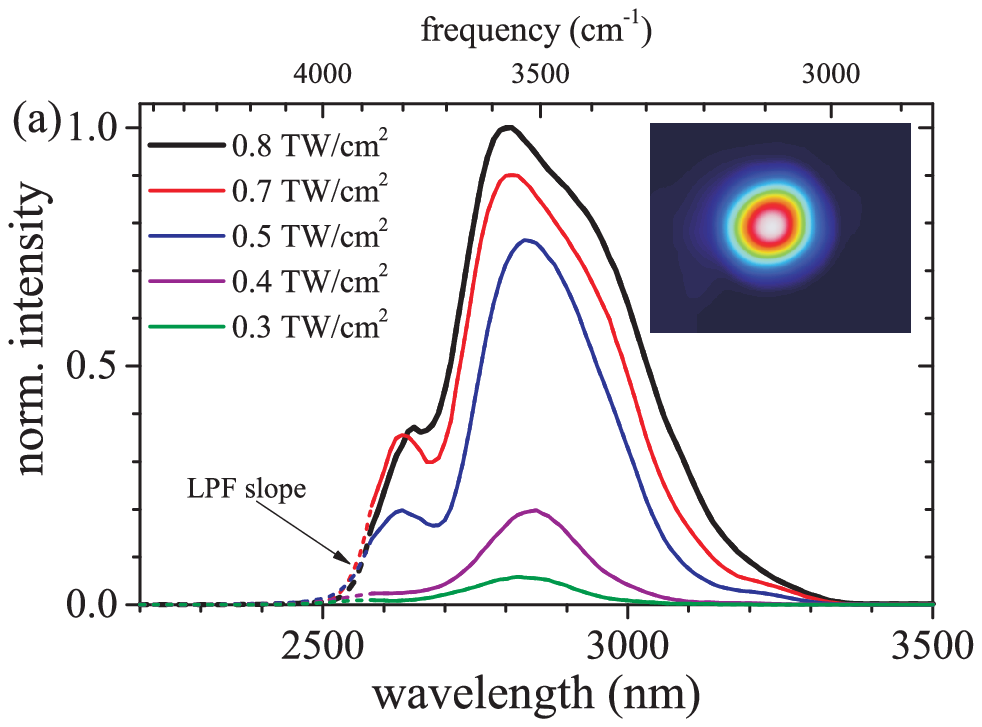}}
\fbox{  \includegraphics[width=0.9\linewidth]{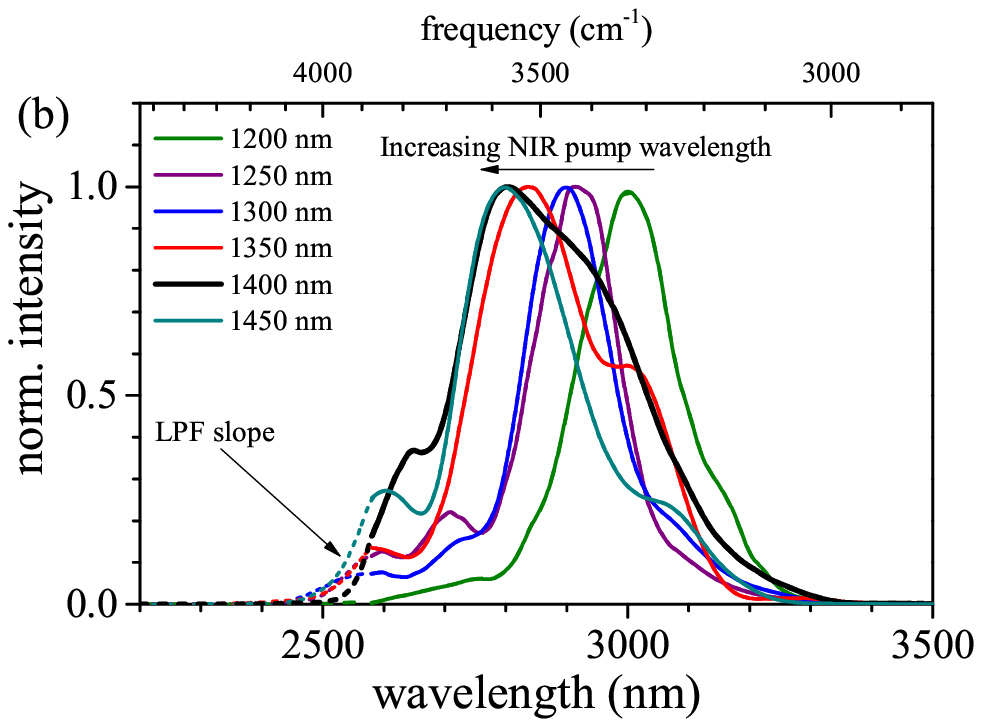}}
  \caption{
  Filtered MIR spectra (linear scale) using a long-pass filter (cut-on wavelength 2400 nm). (a) Evolution of the MIR spectrum for $\lambda_1=1.4\mic$ and sweeping the pump intensity. The spectra are recorded under similar conditions and are normalized to the peak intensity of the $0.8~\rm TW/cm^2$ case. Inset: beam profile of the filtered MIR pulse at $0.8~\rm TW/cm^2$. (b)  Normalized spectra recorded under various NIR pump wavelengths and using the maximum intensity available (see Table \ref{tab:exp}).
  }\label{fig:spectra1}
\end{figure}

Fig. \ref{fig:spectra} shows our experimental results obtained with a simpler technique where a bulk quadratic nonlinear crystal is pumped with a single near-IR (NIR) femtosecond pulse, and if the pump pulse is intense enough to excite a NIR soliton an energetic MIR femtosecond pulse is generated as a so-called dispersive wave (DW) phase-matched to the soliton. The scheme relies on the excitation of a self-defocusing soliton, which here is done in a standard bulk quadratic $\chi^{(2)}$ nonlinear crystal pumped in the normal group-velocity dispersion (GVD) regime. Its simple pumping scheme and high efficiency promise well compared to the traditional approaches, and as filamentation is suppressed because of the self-defocusing interaction, the scheme can be scaled arbitrarily in energy. Fig. \ref{fig:spectra} shows a typical supercontinuum ($\sim 1.5$ octaves), and part of this is a MIR DW in the $3.0\mic$ region, having app. 1,000 cm$^{-1}$ of bandwidth and with a tail extending all the way to $4.0\mic$. It was generated in an unpoled lithium niobate (LiNbO$_3$, LN) crystal, cut for noncritical strongly phase-mismatched "cascaded" second-harmonic generation (SHG), which supports self-defocusing solitons in the pump regime we investigate, $\lambda_1=1.2-1.45~\mu\rm m$ \cite{2012Zhou}. The NIR soliton will then emit a phase-matched DW in the MIR \cite{2011Bache}, where LN has anomalous GVD ($\lambda>\lambda_{\rm ZD}$, where the zero-dispersion wavelength is $\lambda_{\rm ZD}=1.92\mic$). Experimentally we found that the DW shows up for a range of pump intensities, see Fig. \ref{fig:spectra1}(a) where the evolution for $\lambda_1=1.4~\mu\rm m$ is shown, and the filtered DW has a near-Gaussian transverse beam profile. We also found that the DW is tunable over 200 nm by sweeping the pump wavelength, see Fig. \ref{fig:spectra1}(b). The experimental realization of NIR self-defocusing soliton self-compression and supercontinuum generation in a bulk LN crystal cut for noncritical interaction was first demonstrated by some of us \cite{2012Zhou}. Self-defocusing solitons have also been observed experimentally using critical SHG, namely in periodically poled LN \cite{2004Ashihara,2008Zeng}, and in beta-barium borate ($\beta$-BaB$_2$O$_4$) \cite{2002Ashihara,2006Moses,2014Zhou}, while supercontinuum generation from cascaded SHG in periodically-poled LN waveguides has also been observed \cite{2007Langrock,2011Phillips}. However, no evidence of a DW has so far been presented. Its existence was predicted by some of us \cite{2008Bache,2010Bache}, where we also suggested it as an efficient source of MIR femtosecond pulses \cite{2011Bache}. Here we provide the experimental confirmation of this. 

Soliton-induced DWs (also denoted "resonant radiation" or "Cherenkov radiation") have been intensively investigated for generating waves at new frequencies, e.g. in fiber supercontinuum generation \cite{2003Skryabin,2006Dudley,2010Skryabin}.
They were also used for wavelength conversion, especially in the short-wavelength side of the soliton (normal dispersion regime) \cite{2003Tartara,2010Chang,2013Mak}.
Inside specially designed photonic crystal fibers with small cores and double zero-dispersion wavelengths, fiber-based DWs located to the red side of the soliton were also reported \cite{2008Falk,2010Kolesik,2013Yuan}, and with a NIR pump, the longest obtained emission wavelength is about $2.2\mic$ \cite{2013Yuan}. One of the major limitations of fiber-based DW generation is the limited pulse energy, especially for DW generation on the red side, which requires fibers with double zero-dispersion wavelengths. Using bulk materials obviously allows to use a much higher pulse energy, but much less effort has been made for this research direction. In bulk, the self-focusing Kerr nonlinearity tends to generate filamentation, and bulk DWs have mainly been observed in connection with conical emission or the characteristic spatio-temporal spectra of the self-focusing filamentation \cite{1990Golub,1996Nibbering,2005Kolesik,2007Faccio}. Another issue is that in bulk self-focusing Kerr media the DW is always located on the blue side of the soliton, such as the UV DWs recently observed \cite{2012Rubino}, and can therefore not be used for downconversion to the MIR. 

Using instead a self-defocusing nonlinearity, 
the soliton formation will be filament free and the DW will be located on the red side of the soliton \cite{2008Bache,2010Bache} allowing for MIR pulse generation \cite{2011Bache}. Such a self-defocusing effect can be generated through phase-mismatched (cascaded) quadratic nonlinear process, e.g. SHG \cite{1992Desalvo}. Under strongly phase-mismatched conditions, the pump experiences a Kerr-like nonlinearity characterized by the cascaded nonlinear refractive index $n_{2, \rm casc}^I\propto -d_{\rm eff}^2/\Delta k$, which is negative when $\Delta k=k_2-2k_1>0$. Here $d_{\rm eff}$ is the effective quadratic nonlinearity.


\begin{table}[tb]
  \centering
\caption{Properties of the MIR DWs from in Fig. \ref{fig:spectra1}(b), including input wavelength $\lambda_1$, input energy $W_1$ and peak input intensity $I_1$. The DW center wavelength $\lambda_{\rm DW}$ is calculated as a weighted average over the filtered MIR spectrum. The MIR conversion efficiency $\eta_{\rm MIR}$ was found by measuring the power of the filtered and unfiltered case. The MIR bandwidth $\Delta \nu$ is shown as FWHM and at $-20$ dB. 
  }\label{tab:exp}
      \begin{tabular}{ccc|cccccc}
      \hline 
    $\lambda_1$ & $W_1$ & $I_1$  & $\lambda_{\rm DW}$ & $\eta_{\rm MIR}$ & $\Delta \nu_{\rm FWHM}$ & $\Delta \nu_{\rm -20 dB}$  \\
    $\mu$m & $\mu$J & $\rm TWcm^{-2}$ & $\mu$m & \% & $\rm cm^{-1}$ & $\rm cm^{-1}$ \\
      \hline
    1.20	& 105 & 0.5 & 3.00	& 1.5 & 220	& 815	\\
    1.25	& 190 & 0.9 & 2.88	& 3.5 & 200	& 985	 \\
    1.30	& 210 & 1.0 & 2.90	& 4.8 & 190	& 1,050	 \\
    1.35	& 195 & 0.9 & 2.88	& 5.8 & 375	& 1,060	\\
    1.40    & 168 & 0.8 & 2.87	& 6.3 & 390	& 965	 \\
    1.45	& 125 & 0.6 & 2.83	& 6.0 & 275	& 990	\\
      \hline
    \end{tabular}
\end{table}

The experimental setup for generating the MIR DW is identical to the one reported in \cite{2012Zhou}. It is rather simple, 
consisting of a NIR pump laser source (here a 1 kHz commercial OPA system), which is loosely collimated and projected to the quadratic nonlinear crystal through the telescope consisting of two curved silver reflectors. The beam  spot size on the crystal was 0.6 mm full-width-at-half-max (FWHM), which is large enough to avoid beam divergence in the short crystal, and the large spot size is also intended to keep diffraction effects to a minimum. The pulse durations of the NIR pumps were around 50 fs and nearly transform limited. A neutral density filter wheel was used in front of the telescope to adjust the pump intensity. For spectral characterization, an InGaAs CCD-based spectrometer is used for the wavelength below 2300 nm. A grating monochromator and HgCdTe (MCT) detector connected with a box-car integrator was used to record the MIR spectrum.

Just as in \cite{2012Zhou} we used a 10-mm-long, $5\%$ MgO-doped congruent LN ($10\times10~{\rm mm^2}$ aperture, Altechna). The pump pulse and the generated second harmonic (SH) are both $e$-polarized beams with polarization along the vertical optical $Z$ axis. The crystal was Y-cut ($\theta = \pi /2$, $\left | \phi \right | = \pi /2 $) for noncritical (type 0) $ee\rightarrow e$ SHG. (Note that in \cite{2012Zhou} we mistakenly reported the crystal to be X-cut.) This has $\Delta k\simeq +500~\rm mm^{-1}$, resulting in a defocusing cascaded nonlinearity. Traditionally in type 0 cascading, periodic poling would be employed to reduce $\Delta k$ and thereby increase the cascading strength, but we choose not to do this as it would leave the cascaded nonlinearity resonant and narrow-band \cite{2007Bache-nonlocal,2008Bache,2012Zhou}. Despite the large $\Delta k$, the cascaded nonlinearity is still strong because of the large $d_{\rm eff}=d_{33}=25$ pm/V. What ultimately matters is that the self-defocusing cascading nonlinearity is larger than the self-focusing electronic Kerr nonlinearity in the material so the pump effectively experiences a self-defocusing Kerr effect, and an advantage of such a large $\Delta k$ is actually that the SH conversion remains insignificant, typically on the order of a few percent. Note also that the noncritical interaction has zero spatial walk-off (as opposed to the critical SHG scheme in \cite{2004Ashihara,2002Ashihara,2006Moses,2008Zeng,2014Zhou}).

With this setup, significant spectral broadening occurred when gradually increasing the pump intensity. This was enabled by the ultra-broadband nature of the noncritical strongly phase-mismatched SHG process, which gives rise to soliton self-compression to few-cycle duration and subsequently the formation of a supercontinuum \cite{2012Zhou}. Fig. \ref{fig:spectra} shows the measured spectrum under $1.4\mic$ and $0.8~{\rm TW/cm}^2$ pump condition. It shows tremendous spectral extensions in the red side of the pump and goes far beyond the zero-dispersion wavelength. The total bandwidth is around 1.5 octaves at $-20$ dB (spanning 163 THz, ranging from 3,175-8,600 cm$^{-1}$). This exceeds the 1.1 octave that we previously measured with this setup \cite{2012Zhou} (however, note that in \cite{2012Zhou} the MIR part of the spectrum was not measured).

The MIR DWs appear as prominent peaks beyond the envelope dip around $2.5\mic$. We applied a piece of long-pass filter (LPF) with the cut-on wavelength of $2.40\mic$ (Edmund Optics) to filter away the soliton spectrum and harvest the MIR DW. The typical evolution of the filtered MIR DW under various pump intensities is shown in Fig. \ref{fig:spectra1}(a); note that the LPF slope affects the spectrum up to $2.58\mic$, as indicated with dashed lines. The MIR DW first appears at the pump intensity of $0.3~{\rm TW/cm}^2$, and gradually builds up when the pump intensity further increases. At the maximum pump intensity the MIR bandwidth at $-20$ dB is $965~\rm cm^{-1}$, and covers the range from 3,000 $\rm cm^{-1}$ to 3,965 $\rm cm^{-1}$.

The spatial quality of the filtered MIR pulses was also measured using an uncooled microbolometer camera. The inset of Fig. \ref{fig:spectra1}(a) shows a typical case recorded for the $\lambda_1=1.4\mic$ and 0.8 TW/cm$^2$ case, evidencing that the generated MIR beam has a nice Gaussian beam profile. It here is worth to mention that even under the maximum pump intensities, the output beam is stable and accompanied with relatively pure visible color from SHG. This confirms that the total nonlinearity is self-defocusing, enabling a filament-free supercontinuum.


Fig. \ref{fig:spectra1}(b) shows the filtered DWs found with various pump wavelengths by tuning the OPA wavelength from $1.20-1.45\mic$ with 50 nm steps, and using the maximum input intensity available (see Table \ref{tab:exp}). All these wavelengths lie in the normal dispersion regime for the LN crystal, which has $\lambda_{\rm ZD}=1.92 \mic$. 
The so-called effective soliton order \cite{2007Bache} was calculated to be from 1.5-4.0 under the various pump wavelengths used, and the intensities were therefore high enough to excite solitons.
A MIR peak centered at $3.00\mic$ was obtained with the $1.20\mic$ pump and a MIR peak centered at $2.83\mic$ was found from $1.45\mic$ pump. This shows a tunability over nearly 200 nm, and this is the expected trend that when the pump wavelength decreases, the phase-matching wavelength of the MIR pulses increases. Specifically the DW phase-matching condition is $k_{\rm sol}(\omega)=k_{\rm DW}(\omega)$, where $k_{\rm sol}(\omega)=k_1(\omega_{\rm sol})+(\omega-\omega_{\rm sol})/v_{g,\rm sol}+q_{\rm sol}$ is the soliton wavenumber and $k_{\rm DW}(\omega)=k_1(\omega)$ is the DW wavenumber, which simply follows the material dispersion, in this case $k_1(\omega)$. The dispersion-free nature of the soliton is reflected in its linear dependence in frequency, which simply states that it is a wave-packet traveling with the group velocity $v_{g,\rm sol}$. There is also a nonlinear phase contribution to the soliton phase \cite{1995Akhmediev} $q_{\rm sol}=n_{2,\rm eff}^II_{\rm sol} \omega_{\rm sol}/(2c)$ where $n_{2,\rm eff}^I=n_{2,\rm casc}^I+n_{2,\rm Kerr}^I$ is the effective nonlinear index and $n_{2,\rm Kerr}^I$ is the Kerr electronic nonlinear index \cite{2010Bache}. For an effective defocusing nonlinearity ($n_{2,\rm eff}^I<0$) this nonlinear contribution will shift the phase-matching condition slightly towards longer wavelengths. However, for most cases this shift is insignificant compared to the impact of shifting the NIR soliton wavelength; thus, the tunability of the MIR DW relies mainly on tuning the NIR soliton wavelength, and the phase-matching curve for LN with $q_{\rm sol}=0$ can be appreciated in Fig. 1(a) in Ref. \cite{2011Bache}.


However, as the pump wavelength was varied the wavelength tunability of the MIR DWs was weaker than expected from theory. According to Fig. 1(a) in Ref. \cite{2011Bache} a pump soliton in the $1.20-1.45~\mu$m range should namely emit DWs with center wavelengths $\lambda_{\rm DW}\sim 3.0-5.0~\mu$m, but instead we find $\lambda_{\rm DW}\sim 2.8-3.0~\mu$m. Part of the explanation is to be found in the Raman-induced soliton self-frequency shift: LN has been found to have a significant Raman nonlinearity (see review in \cite{2012Bache}), and the NIR soliton therefore will be significantly red-shifted compared to the input wavelength. This explains why DW radiation is found at lower wavelengths than expected a priori from the input wavelength. However, this alone cannot explain the results from Fig. \ref{fig:spectra1}. 


\begin{figure}[tb]
  \centering
\fbox{  \includegraphics[width=\linewidth]{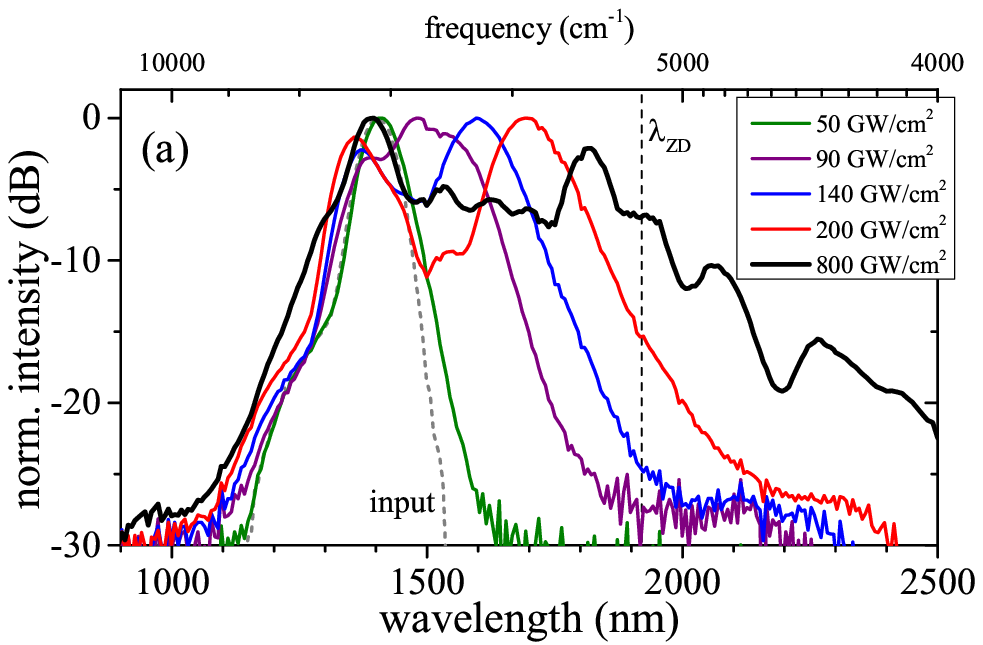}}
\fbox{  \includegraphics[width=\linewidth]{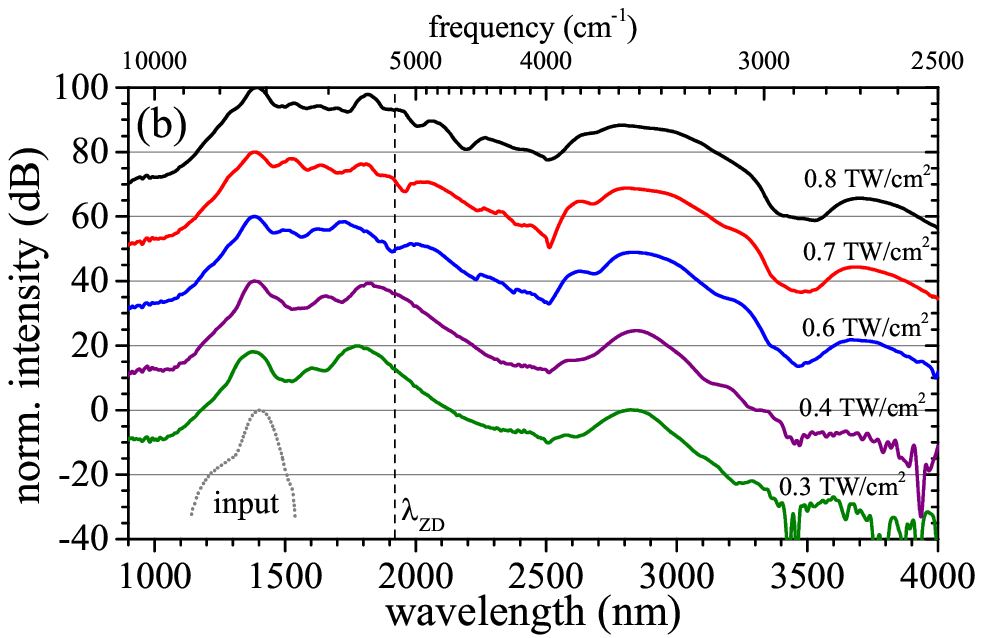}}
\caption{Variation of the supercontinuum content at $\lambda_1=1.4\mic$ while sweeping the input peak intensity. (a) The NIR development at low intensities (where no MIR spectral content was measurable); for comparison the spectrum for the maximum intensity is also shown. (b) The full spectra at high intensities (note that a 20 dB offset per curve is used for clarity of presentation).
}\label{fig:scg}
\end{figure}

In order to investigate this further, we show first in Fig. \ref{fig:scg} how the spectral broadening develops when increasing the intensity. Fig. \ref{fig:scg}(a) shows the low-intensity NIR development, and significant broadening starts to occur between 50 and 90 GW/cm$^2$, after which the new spectral shoulder clearly redshifts with increased intensity. This indicates that the Raman effect is affecting the dynamics; as discussed in \cite{2013Guo} the FW spectral phase is affected by competing cascading and Raman effects, and when the Raman term is strong the spectrum will red-shift. Fig. \ref{fig:scg}(b) shows the high-intensity development across the entire NIR and MIR; with these high intensities the MIR part of the spectrum started to appear. We note there are two DW peaks, since besides the main one just below $2.9\mic$ there is also a minor peak between $3.5-4.0\mic$. The presence of this minor peak leads us to believe that it is an early-stage DW emitted at the self-compression point when the soliton first forms. The low-wavelength major peak would then be a DW emitted at a later stage. The reason why the low-wavelength DW is dominating in the spectrum can then be explained by the Raman red-shifting effect: at the initial self-compression point the soliton has only been red-shifted slightly, and has therefore a phase-matched DW far into the MIR. This DW will therefore have a weak coupling to the soliton and thereby a low radiation efficiency. As the soliton propagates further it relaxes and recompresses, where it again will emit a dispersive wave. During this stage the Raman effect will red-shift the soliton further, which leads to a phase-matching condition at a lower MIR wavelength, and hence the coupling to the soliton is much stronger leading to a higher MIR DW radiation efficiency. As the soliton comes closer to the zero-dispersion wavelength it will be spectrally recoiled so its redshifting will be arrested, and therefore the blue-shifting of the DW emission will also arrest.

Going back to the low-intensity dynamics of Fig. \ref{fig:scg}(a), we note that the spectrum is substantially broadened and red-shifted but the MIR DW radiation was not observed for these cases. For sure the lowest intensities are not enough to form a soliton, but as the intensity is increased the soliton onset will occur at some point. It is here relevant to point out that once the soliton formation threshold is crossed, it is still required to see a significant spectral broadening in order to observe a MIR DW, i.e. a higher-order soliton must be excited. Once this happens the initial soliton self-compression point must occur within the relatively short crystal length we have chosen. Considering this, 200 GW/cm$^2$ might be enough for soliton self-compression to occur within 10 mm, and thereby also to form a DW. However, even if a soliton has formed already at this level, the coupling between the soliton and the DW is evidently not high enough for observing the MIR radiation. This would be in line with what we typically see in simulations: at the soliton self-compression point the DW radiation is very weak, merely a bump on the soliton spectral tail in the anomalous dispersion regime (see, e.g., \cite{2011Bache}). At 300 GW/cm$^2$ we do observe significant MIR radiation in form of DWs, and notice that the spectral shoulders of the soliton in the normal dispersion regime are located quite close to the zero-dispersion wavelength. This happens due to the Raman self-frequency shift. Usually a further increase in the intensity would lead to a stronger Raman red shift of the soliton spectrum, but the spectral recoil of the anomalous dispersion regime leads to a saturation of the red-shift: clearly the higher-intensity spectra do not present any change in the red-shift. This in turn leads to DWs that are emitted at roughly the same MIR wavelength, independent of the intensity. Therefore we conclude that the reason why the center wavelength of the major DW remains fixed while sweeping the intensity, as in Fig. \ref{fig:spectra1}(a), is because the Raman self-frequency shift leads to a significant red-shift of the spectrum, so that at the onset of soliton-formation in the 10 mm crystal the red-shift is already close to saturation caused by the spectral recoil of the anomalous dispersion regime.

\begin{figure}[tb]
  \centering
\fbox{  \includegraphics[width=\linewidth]{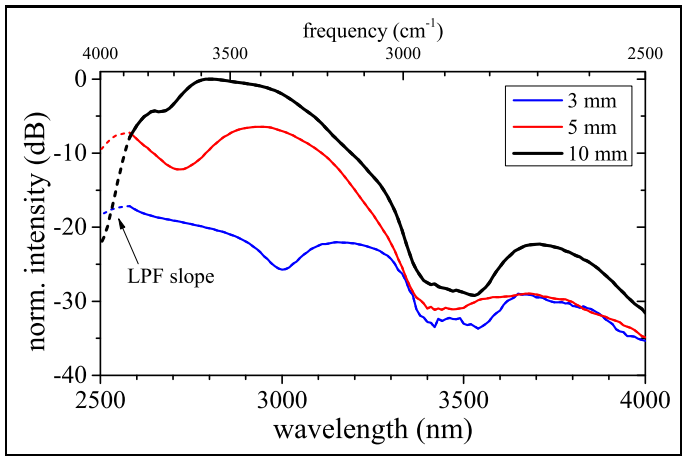}}
\caption{Variation of the LPF MIR spectrum for 3, 5 and 10 mm LN and using $\lambda_1=1.4\mic$ and $I_1=0.8~\rm TW/cm^2$.
}\label{fig:LN-length}
\end{figure}

In order to support these claims, we investigate the early-stage DW dynamics, and Fig. \ref{fig:LN-length} shows the LPF MIR spectrum using 3 and 5 mm crystals, and they are compared to the 10 mm case we have already presented above. The 3 and 5 mm spectra have been normalized relative to the 10 mm case so that the areas under the curves roughly reflect the amount of energy we estimate the MIR spectra contained after the LPF. This approach is justified by the excellent spatial quality of the filtered MIR spectrum, see inset in Fig. \ref{fig:spectra1}(a). It is clear that all three cases have the minor peak, which does not change position with propagation although it is more powerful at 10 mm. However, the major peak grows significantly upon propagation and shifts its center wavelength from $\sim 3.2\mic$ to $\sim 2.9\mic$. This supports the idea that Raman effect is red-shifting the soliton so that the DW radiation will become blue-shifted during propagation, and that several DWs are emitted.

The Raman self-frequency shift can in a similar way also explain the fact that in the entire pump range from $\lambda_1=1.20-1.45\mic$ we observe a (main) DW with a relatively limited tuning range from $2.8-3.0\mic$. For low $\lambda_1$ the early-stage DW will be extremely weak because the phase-matching wavelength lies quite far away from the soliton. However, since the Raman effect will then shift the soliton we are able to observe a strong DW nonetheless, formed later in the crystal, but it is much more blue-shifted than what one would expect from the input wavelength as the soliton is now significantly red-shifted. Moreover, because the soliton red-shift will saturate, the low-wavelength input cases will not be so different from the high-wavelength input cases. That being said, we emphasize that there is a degree of tunability in the MIR pulses through the input wavelength, as Fig. \ref{fig:spectra1}(b) demonstrates, and we expect that by increasing the input wavelength further will extend the DW tunability to wavelengths lower than $2.8\mic$. It is also worth to note that LN is well-known to have a relatively strong Raman nonlinearity, so we expect that with other crystals with a less dominating Raman nonlinearity will show stronger tunability concerning the DW wavelength.

The average power of the filtered MIR pulse was measured to 10.5 mW with 168 mW pump power at $1.40\mic$, corresponding to an overall NIR to MIR energy conversion efficiency of $\eta_{\rm MIR}=6.3\%$. We generally find $\eta_{\rm MIR}=4-6$\% (see Table \ref{tab:exp}), except for $\lambda_1=1.20\mic$, where it is $\eta_{\rm MIR}=1.5\%$. We expected this case to have the lowest efficiency because it had $\lambda_1$ furthest away from $\lambda_{\rm ZD}$ and thereby also the phase-matching point further into the MIR than the other cases; this naturally gives a lower coupling efficiency. Moreover, the lower OPA output gave a significantly lower peak input intensity than the other cases, and as the GVD increases at lower wavelengths this all leads to a lower soliton order and thus a reduced spectral broadening. These factors combined are behind the lower MIR conversion efficiency at $1.20\mic$. 

Because we exploit a noncritical cascaded SHG interaction, the soliton formation and MIR DW generation is rather insensitive to the crystal angle adjustment, which is quite different from the critical phase-matching condition required by other techniques like OPA. 
Moreover, the conversion efficiency is already comparable with the traditional DFG process under critical phase-matching conditions and synchronized pump wavelengths. We could actually expect much higher conversion efficiencies when pumping closer to the zero-dispersion wavelength (see \cite{2011Bache}), with the tradeoff that the DW will be generated closer to the pump wavelength as well. This is because the DW energy is proportional to the soliton spectral overlap at the radiated frequency, so the closer we pump to the zero-dispersion wavelength, the more energy will reside in the DW.

We used few-GW peak-power pump pulses as this was the limit of our OPA, but our scheme also supports higher peak powers because it is filament-free due to the self-defocusing nonlinearity. This gives a strong scalability with input energy. Consider, e.g., using a TW-peak power Cr:forsterite laser amplifier (sub-100 fs $>$100 mJ pulses at 1250 nm), and with 3\% efficiency this could generate around 3 mJ MIR energy in a 10 mm long LN crystal provided that the crystal aperture is increased to around $40\times 40~\rm mm^2$. Such a large aperture is only possible because periodic poling is not needed. The MIR spectral density would be close to 1 mW/nm at a 10 Hz repetition rate.
\begin{figure}[tb]
  \centering
\fbox{    \includegraphics[width=\linewidth]{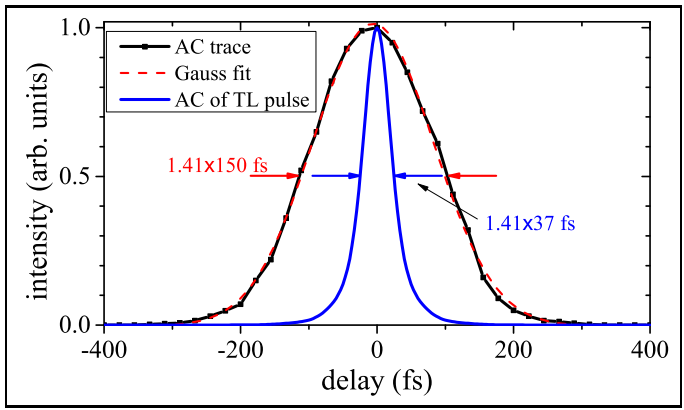}}
\caption{Typical intensity autocorrelation trace for the filtered MIR pulses generated by a 1.3 $\mu $m pump with $I_1=1.0~\rm TW/cm^2$. The blue curve shows the transform-limited pulse duration of the filtered MIR spectrum. 
}\label{fig:AC}
\end{figure}

We temporally characterized the MIR DWs with a home-made MIR SHG intensity autocorrelator, consisting of a ZnSe beam splitter and a 0.4-mm-thick AgGaS$_{\rm2}$ crystal ($\theta=39^{\circ}$, $ \phi=45^{\circ}$, Eksma Optics). 
A typical measurement for $\lambda_1=1.30\mic$ is shown in Fig. \ref{fig:AC}. The pulse duration is 150 fs assuming a Gaussian shape. Similarly, for $\lambda_1=1.40\mic$ we obtained a 190 fs pulse duration, and between 145 to 210 fs under the other pump wavelengths. Considering the observed MIR bandwidths, this indicates that all the MIR pulses are quite chirped. This is expected as the soliton self-compression point occurs quite early inside the crystal, and after this the DW accumulates phase according to the material dispersion (see also \cite{2011Bache}). We expect the accumulated chirp to be quite linear as GVD is the dominating dispersion at the MIR wavelength. Therefore significant pulse compression is possible by compensating the linear chirp with a dispersive element. This is interesting since the MIR spectra typically support pulse duration around as low as 3-4 optical cycles; in the figure we also show the autocorrelation trace for a transform-limited pulse calculated from the isolated MIR spectrum, and $\simeq 40$ fs duration at $3.0~\mu$m wavelength corresponds to 4 optical cycles. To support this idea, we saw significant pulse compression in realistic numerical simulations where the MIR part was isolated as in the experiment and simple 2. order dispersion compensation was employed.

Concluding, we experimentally observed high-energy efficient and broadband femtosecond MIR wave generation using a NIR femtosecond pump and a short bulk quadratic nonlinear crystal.
These MIR waves were generated by exciting self-defocusing NIR solitons, induced by a cascaded quadratic $\chi^{(2)}$ nonlinearity, and these solitons emitted DWs in the anomalous GVD regime in the MIR. 
Broadband, femtosecond MIR pulses tunable from $2.8-3.0\mic$ were obtained through a 10-mm-long bulk lithium niobate crystal and the tunability can be expanded by pumping over a broader wavelength range. The largest MIR bandwidth generated was almost 1,100 cm$^{-1}$, and covered the range from $3,000-4,100$ cm$^{-1}$. The pulse durations for the generated MIR pulses were measured to be $\sim 145-210$ fs FWHM. This indicates an additional spectral phase that we expect mainly comes from dispersive propagation in the crystal, so obtaining much shorter few-cycle pulses should be possible after simple dispersion compensation. 
Weaker MIR DW emission was found up to $4.0\mic$ (2,500 cm$^{-1}$), and by investigating the spectral contents when sweeping pump wavelength and input intensity as well as various crystal lengths we found that the soliton dynamics and thereby also DW formation seem strongly influenced by Raman self-frequency shifting of the soliton.
The most energetic MIR pulse contained 10.5$~\mu$J of energy, corresponding to $6.3\%$ NIR-to-MIR conversion efficiency. This is comparable to the commonly used OPA/DFG techniques for high-energy MIR pulse generation, and also in other features does the demonstrated method compete well as it shows great simplicity: it does not require synchronized pump wavelengths or critical phase-matching conditions, and it should scale linearly with an increased pump pulse energy. Its simplicity is further emphasized by the fact that no periodic poling is required, implying that large-aperture low-cost and low-complexity crystals can be used. This paves way for a wide range of NIR and MIR nonlinear crystals with large diagonal tensor nonlinearities, which could be used to realize coverage in various MIR wavelength regimes. One specific example is the biaxial LiInS$_{2}$ crystal cut for noncritical interaction, which when pumped in the short edge of the MIR should give a DW in well into the MIR \cite{2013Bache}. This would target a different MIR regime where carbon double and triple bonds have characteristic resonance frequencies and where ultrafast spectroscopy of water is performed to study the H-O-H bending dynamics. We hope this work could pave the way for a new kind of practical ultrafast MIR source, which could be a important complement for the OPA/DFG MIR sources.


\section*{Funding Information}
The Danish Council for Independent Research (11-106702).

\end{document}